# High-Q photonic crystal Fabry–Pérot micro-resonator in thin-film lithium niobate

Hyeon Hwang,[1,a] Seokjoo Go,[1] Guhwan Kim,[2] Hong-Seok Kim,[2] Kiwon Moon,[2] Jung Jin Ju,[2] Hansuek Lee,[1] and Min-Kyo Seo [1,b]

[1]Department of Physics, Korea Advanced Institute of Science and Technology (KAIST), Daejeon, 34141, Rep. of Korea

[2]Electronics and Telecommunication Research Institute (ETRI), Daejeon, 34129, Rep. of Korea

a) e-mail: hyeon_hwang@kaist.ac.kr

b) e-mail: minkyo_seo@kaist.ac.kr

**Abstract:**

Thin-film lithium niobate (TFLN) has emerged as a powerful platform for integrated nonlinear and quantum photonics, owing to its strong optical nonlinearities, wide transparency window, and electro- and piezo-optic properties. However, conventional traveling-wave resonators—such as micro-rings, disks, and racetracks—suffer from curvature-dependent group dispersion and losses, limited spectral tunability, and parasitic nonlinearities, which constrain their performance, scalability, and operational stability in nonlinear photonic circuits. Here, we present photonic crystal (PhC) Fabry–Pérot (FP) micro-resonators in TFLN that address these limitations. The device features a one-dimensional straight cavity bounded by PhC reflectors and supports well-confined standing-wave resonant modes within an engineered photonic bandgap. We achieve intrinsic quality (Q) factors of up to $1.4 \times 10^6$ and demonstrate that both the free spectral range (FSR) and coupling strength can be consistently controlled via cavity length and PhC coupler design, respectively. The photonic bandgap is tunable across the S-, C-, and L-bands without degradation of resonator performance. Spectral confinement of high-Q resonant modes is expected to mitigate parasitic nonlinearities, such as Raman scattering. These advances, together with the one-dimensional geometry, establish PhC FP micro-resonators as compact and scalable building blocks for high-density photonic integrated circuits targeting next-generation nonlinear and quantum applications.

# 1. Introduction

Lithium niobate (LN) is a key material in integrated photonics,[1–4] owing to its strong electro-optic response and second- and third-order nonlinear optical properties, which makes it essential for high-speed modulation,[2,5–10] frequency comb generation,[11–14] and quantum photonic applications.[15–20] The development of thin-film lithium niobate (TFLN) has further expanded its potential by enabling low-loss, tightly confined optical modes, leading to improved nonlinear efficiency and denser photonic integration.[3,13,14,21–23] Despite these advantages, conventional traveling-wave resonators such as disk, ring, and racetrack resonators face fundamental limitations that hinder further performance improvements. Their reliance on curved waveguides introduces bending losses that degrade quality (Q) factors, which limit device miniaturization.[24–27] In addition, the curvature-dependent group dispersion and birefringence not only constrain flexible tuning of the free spectral range (FSR) but also lead to undesired mode conversions.[27] TFLN traveling-wave resonators support high-Q resonant modes over a very broad spectral range; however, this also enhances Raman scattering and other parasitic nonlinearities outside the frequency range of interest, leading to excess noise, increased power consumption, and instability in second-order nonlinear processes and their device-level implementations.[14,17,28,29] It is therefore timely to develop an alternative micro-resonator architecture that addresses these limitations in TFLN and advances the performance of second-order nonlinear optical processes.

Photonic crystal (PhC) Fabry-Pérot (FP) micro-resonators offer a promising alternative to traveling-wave resonators. The standing-wave resonant modes formed in a straight cavity between two PhC reflectors inherently eliminate bending losses and curvature-dependent dispersion,[30,31] enabling not only high Q-factors but also reliable, length-dependent FSR control. In silicon nitride ($Si_3N_4$) platforms, PhC FP micro-resonators have been utilized to advance third-order nonlinear effects and their applications, including low-threshold optical parametric oscillation, soliton microcomb formation, and high-precision injection locking.[30,32–35] The deterministic support of high-Q resonant modes within the photonic bandgap is anticipated to effectively suppress parasitic nonlinearities outside its spectral range. Furthermore, the one-dimensional geometry of FP micro-resonators is suited to facilitate high-density on-chip integration.[30,34]

In this work, we extend the concept of PhC FP micro-resonators to TFLN, aiming to harness its strong second- and third-order nonlinearities, electro-optic tunability, and low-loss waveguiding capabilities. By integrating PhC reflectors into a one-dimensional TFLN waveguide platform, we realize a high-Q micro-resonator architecture that achieves a record intrinsic Q-factor of up to $1.4 \times 10^6$. The PhC reflectors support high-Q standing-wave resonant modes over an engineered photonic bandgap, providing strong mode selectivity and spectral confinement across the telecommunication wavelength range. The FSR of the transverse-electric (TE) standing-wave resonant modes scales linearly with the cavity length while maintaining a constant group index, offering greater design flexibility for applications requiring precise spectral control. We systematically extract the propagation loss in the long waveguide (0.27 dB/cm) and the interfacial loss at the PhC reflector (0.07 dB), which together determine the intrinsic Q-factor, and establish a straightforward framework for controlling the coupling Q-factor via the PhC coupler to optimize resonator-waveguide energy exchange. By varying the period of the PhC lattice, the photonic bandgap is tuned to not only cover the C-band but also extend into the S- and L-bands. These results position the PhC FP micro-resonator as a highly scalable, compact, and versatile component for next-generation integrated nonlinear and quantum photonic devices.

## 2. Schematic and structural design of the high-Q PhC FP micro-resonator

The high-Q PhC FP micro-resonator comprises three main components: a PhC coupler ((1) in Fig.1 (a)), a central waveguide ((2) in Fig.1 (a)), and a high-reflection PhC mirror ((3) in Fig.1 (a)). The PhC coupler controls the coupling strength by adjusting the number of unit cells, regulating the amount of light entering and exiting the cavity. The central waveguide serves as the optical cavity, where its width gradually transitions from a narrow width at the PhC coupler interface to a wider width at the center before tapering back to a narrow width at the perfect reflector interface. This tapered structure minimizes scattering loss induced by sidewall roughness while maintaining strong optical confinement. The perfect reflector, designed with a sufficiently large number of PhC unit cells, ensures near-total reflection of the transmitted light back into the cavity, thereby forming a standing wave between the coupler and perfect reflector within the photonic bandgap. Figure 1(b) presents an optical microscope image of the fabricated high-Q PhC FP micro-resonators.

The photonic crystal reflector is designed to achieve a wide photonic bandgap, enabling the support of a large number of high-Q resonant modes, which are promising for future applications in Kerr and Electro-Optic frequency combs. The PhC reflector, as shown in Fig. 1(c), features a fishbone-shaped structure with a period ($a$), a central waveguide width ($W_2$), and a corrugation width ($W_c$). A one-dimensional PhC lattice with $a$ = 435 nm, $W_2$ = 250 nm, and $W_c$ = 700 nm opens a wide bandgap of over ~50 nm in the telecommunication wavelength range for the fundamental transverse-electric (TE) guided mode, as will be discussed in detail later.

A key challenge in achieving a high-Q factor is minimizing mode conversion loss at the PhC reflector interface. This is addressed by implementing an apodization method, which adiabatically transitions $W_2$ and $W_c$ to mitigate mode mismatch between the propagating wave in the coupled waveguide and cavity and the Bloch wave of the PhC. Moreover, the apodization method suppresses crosstalk with leaky modes. The detailed design method is described in Method A. The central waveguide width of the PhC reflector in Fig. 1(c) sinusoidally transitions from the width of the coupled waveguide ($W_1$) of 700 nm to its final value over 30 unit cells, while the initial corrugation width ($W_c$) increases from 0 nm.

The fabricated PhC micro-resonator, as shown in Fig. 1(b), has a significantly smaller footprint compared to conventional micro-ring resonators. We find that this compact design is advantageous for photonic integration, allowing higher device density without compromising performance. The samples were fabricated by electron beam lithography on FOX-16 resist (Dow Corning), Ar-based ICP-RIE etching, and $H_2O_2$ + ammonia-based wet etching, as described in detail in the following paper.[10,20,36,37] Figure 1(d) shows the cross-sectional structure of our device, where the 450 nm thick LN film was etched down by 210 nm, with a sidewall angle of 75 degrees.

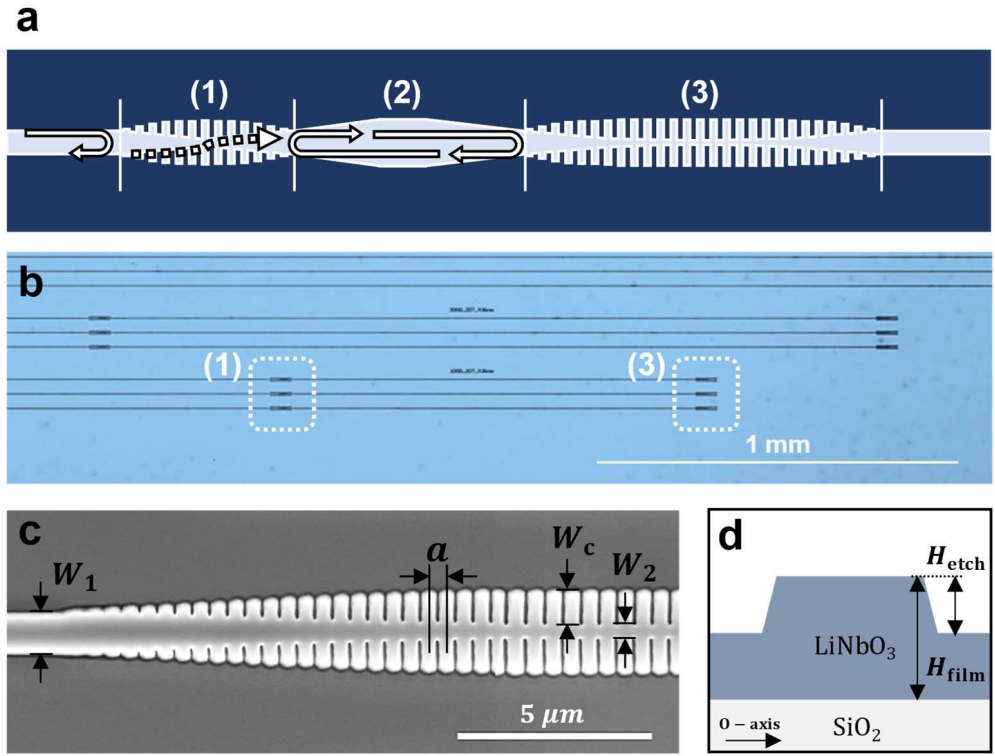

**Fig. 1** (a) Schematic illustration of the high-Q PhC FP micro-resonator, consisting of a PhC coupler (1), a central waveguide cavity (2), and a PhC perfect reflector (3). (b) Optical microscope image of the fabricated PhC micro-resonators. The white boxes indicate the PhC reflectors. (c) Scanning electron microscope (SEM) image of the PhC reflector. $W_1$ = 700 nm, $W_2$ = 250 nm, $W_c$ = 700 nm, and $a$ = 435 nm. (d) Cross-sectional schematic of the waveguide and cavity. $H_{etch}$ = 210 nm and $H_{film}$ = 450 nm. The sidewall angle is 75 degrees.

## 3. Characterization of resonant modes and Q-factors

To characterize the resonant modes of the PhC FP micro-resonator, a reflectance measurement setup was employed, as illustrated in Fig. 2(a). The interference between the partially reflected light from the coupler and the out-coupled light of the resonant mode circulating inside the cavity results in a Lorentzian-shaped reflectance dip, similar to that observed in typical transmission measurements of micro-ring resonators. A representative reflection spectrum measured from an under-coupled cavity with a length of 4 mm is shown in Fig. 2(b). High-Q resonant modes with a narrow linewidth of ~200 MHz (inset of Fig. 2(b)) appear within the 1540–1580 nm wavelength range. The overall reflectance envelope decreases outside the photonic bandgap, corresponding to regions of increased optical transmission through the entire device. We note that the variation in the resonance depth across different modes (black dots in Fig. 2(b)) arises from mode crossings with low-Q modes induced by partial reflections at the cleaved end facet of the chip. These parasitic reflections can be significantly suppressed by employing inverse taper couplers or adiabatic taper couplers.

Figure 2(c) presents the extracted loaded ($Q_{load}$), intrinsic ($Q_{int}$), and coupling ($Q_{cpl}$) quality factors for all resonant modes, reflecting the under-coupled condition. Within the photonic bandgap, the intrinsic Q-factor exhibits a relatively uniform distribution with a mean value of $9.8 \times 10^5$. The inset of Fig. 2(c) shows the statistical distribution of the intrinsic Q-factor for the ~300 resonant modes in the wavelength range of 1540 to 1580 nm. Outside the bandgap, both the intrinsic and coupling Q-factors decline, which correlates with a reduction in the reflectance of the PhC reflector and coupler. In the following sections, we also examine the dependencies of the Q-factors on the cavity length, photonic band engineering, and coupler design to point out advantages of the PhC FP micro-resonator platform.

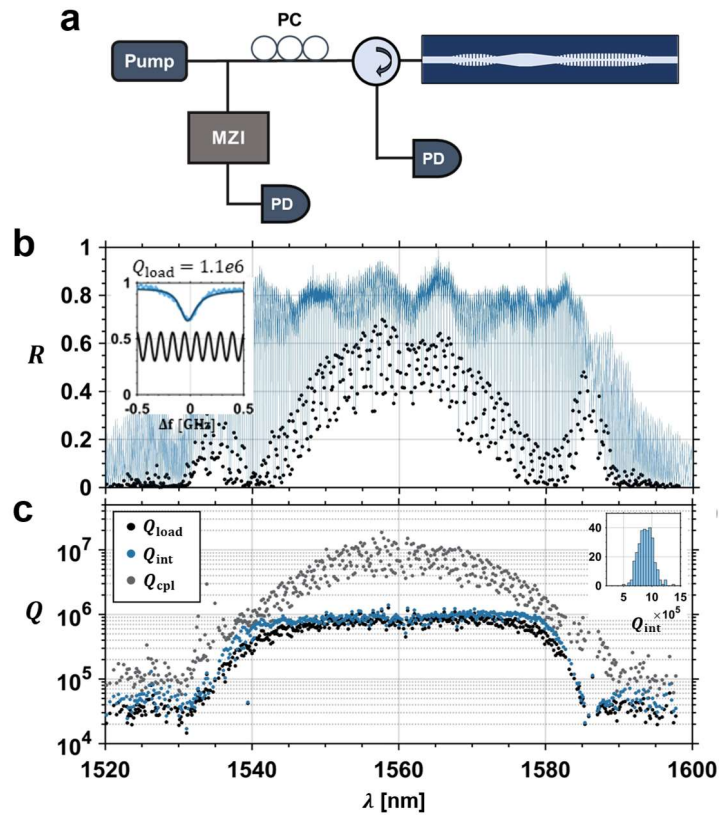

**Fig. 2** (a) Schematic of the reflectance spectrum measurement setup. PC: Polarization controller, PD: photo-detector, MZI: Mach-Zehnder interferometer (b) Measured reflection spectrum for an under-coupled PhC FP micro-resonator with a cavity length of 4 mm. The black dots correspond to the central dips of the resonances. Inset: a representative spectrum of the resonant reflection (blue) and the MZI signal with a 112 MHz bandwidth (black). (c) Extracted loaded ($Q_{load}$), intrinsic ($Q_{int}$), and coupling ($Q_{cpl}$) quality factors for each resonant mode (black dots in (b)). Inset: histogram of $Q_{int}$ for the resonant modes in the wavelength range of 1540-1580 nm.

## 4. Free spectral range scaling and Q-factor characterization

We investigate the free spectral range (FSR) and group index ($n_g$) of the fundamental TE modes in the one-dimensional PhC FP micro-resonator by varying its cavity length ($L$). The PhC FP micro-resonator enables precise FSR tuning simply by adjusting the cavity length. Figure 3(a) presents the measured FSR in PhC FP micro-resonators with cavity lengths of 1, 2, 4, and 7.8 mm (blue dots). The FSR values, determined by averaging the spacing of the resonant modes within the photonic bandgap, exhibit the expected inverse relationship with cavity length:

$$\text{FSR} = \frac{C}{2n_g L + \delta}$$

Where $\delta$ is the additional optical path length by tapering part and partial penetration into PhC reflector. The group index ($n_g$ = 2.24), calculated using the finite-difference frequency-domain (FDFD) mode solver (Tidy3D, Flexcompute), provides the theoretical FSR (black solid line in Fig. 3(a)), which accurately reproduces the experimental results. In the simulation, we used the cross-sectional structure shown in Fig. 1(d).

Notably, the PhC FP micro-resonator allows independent scaling of the FSR without altering the group index. In contrast, in a micro-ring resonator, changing its circumference inherently modifies the curvature, which affects both the spatial mode distribution and the group index. This design flexibility, which decouples the group index from cavity length, makes PhC-FP micro-resonators particularly attractive for nonlinear optical applications, such as electro-optic (EO) comb and Kerr frequency comb generation with well-defined repetition rates.[38,39] Notably, in stimulated Brillouin scattering, this tunability enables accurate alignment between the cavity FSR and the narrow Brillouin gain band.[40,41]

We also measured the intrinsic Q-factors of the resonant modes as a function of the cavity length, enabling the development of a theoretical model of the optical loss channels in the PhC FP resonator (Fig. 3(b)). Considering the propagation loss ($\alpha L$) along the cavity and the interfacial loss ($\gamma$) between the PhC mirror and the cavity, we can achieve the following representation of the intrinsic Q-factor[31]

$$Q_{\text{int}} = \frac{2\pi n_g L}{\lambda}(\alpha L + \gamma)^{-1}$$

The interfacial loss is given by the harmonic mean of the losses at the two PhC mirrors. The propagation loss, primarily caused by the scattering due to the sidewall roughness, scales with the cavity length, the interfacial loss remains independent of it. The theoretical model shows excellent agreement with the experimental results, yielding a propagation loss of 0.27 dB/cm and an interfacial loss of 0.07 dB.

The demonstration of a maximum $Q_{\text{int}}$ of $1.4 \times 10^6$ and an average of $0.96 \times 10^6$ for an 8 mm cavity length sets a new benchmark for one-dimensional PhC FP micro-resonators in TFLN.[42–44] The use of a thicker LN layer (450 nm in this work) can further enhance the Q-factor by improving mode confinement, reducing scattering loss, and broadening the photonic bandgap. In addition, similar to how advanced tapering techniques such as Euler bends have been used to mitigate mode conversion losses in bending waveguides, the new apodization methods with optimized functional forms can be used in PhC reflectors to minimize unwanted optical losses while maintaining the desired transmission required for precise coupling strength control.[45]

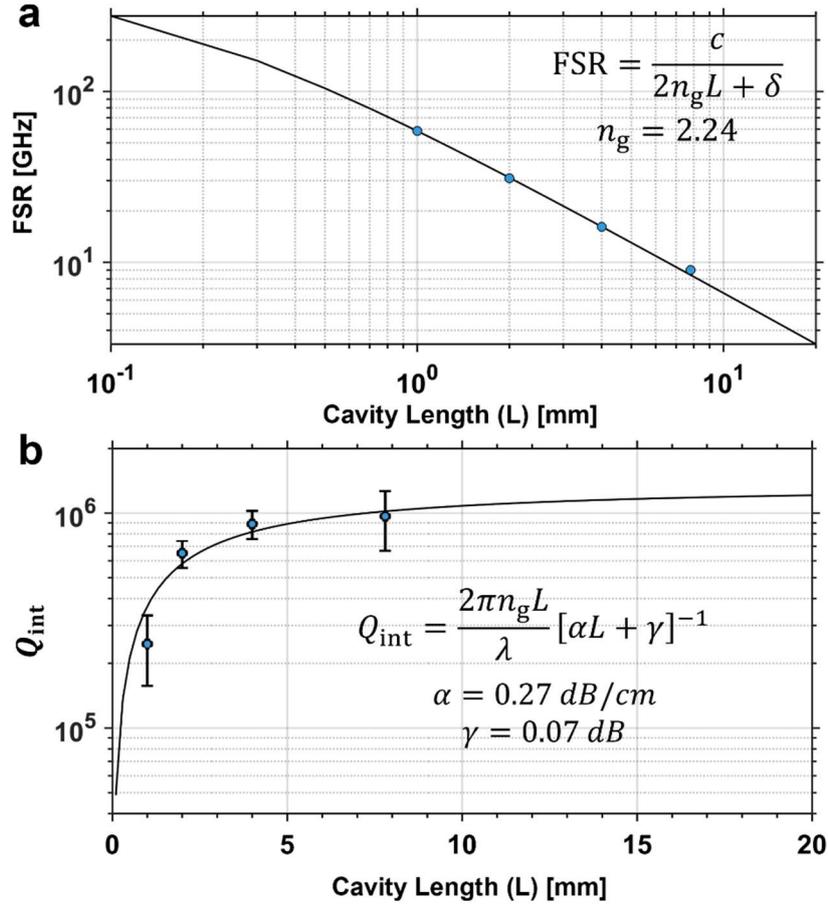

**Fig. 3** (a) Measured free spectral range (FSR) of the fundamental TE modes in PhC FP micro-resonators with cavity lengths of 1, 2, 4, and 7.8 mm (blue dots). The theoretical FSR (black solid line) is calculated using the group index ($n_g$ = 2.24) obtained from finite-difference frequency-domain (FDFD) simulations. (b) Measured (blue dots) and calculated (black solid line) intrinsic Q-factor ($Q_{int}$) as a function of cavity length, illustrating the contributions of propagation loss ($\alpha L$) and interfacial loss ($\gamma$).

## 5. Tunable broadband photonic bandgap

Precise and wide tuning of the operating wavelength range of photonic bands is crucial for tailoring resonant mode's formation and engineering phase-matching conditions in various nonlinear and quantum optical applications. We demonstrate photonic bandgap tuning across an extended telecommunication range, not only covering the C-band but also expanding to the S- and L-bands, by simply varying the period ($a$) of the PhC lattice (Fig. 4). In PhC FP micro-resonators, engineering photonic bands enables the isolated induction and enhancement of nonlinear optical interactions within the bandgap, suppressing parasitic effects, such as Raman scattering by rejecting them outside this spectral range.

Figure 4(a) shows the simulated photonic band structures around the one-dimensional Brillouin zone boundary for PhC periods of 430 nm, 435 nm, and 440 nm. Here, the central waveguide width and the corrugation width of the PhC unit cell are 250 nm and 700 nm, respectively, with a duty cycle of 40% for the corrugation part. As the period decreases, the center frequency of the photonic bandgap blueshifts, while its width remains nearly unchanged. According to the one-dimensional Bragg condition, $\boldsymbol{\lambda_c = 2n_{\text{eff}} \times a}$, the center wavelength of the photonic bandgap scales linearly with the PhC lattice constant when the effective index ($n_{\text{eff}}$) of the Bloch mode does not change significantly.

Figure 4(b) shows the experimentally measured reflection spectra of the PhC FP micro-resonators with different photonic bandgaps. The PhC lattice periods match those used in the simulations. In experiments, the PhC FP micro-resonators, whose mirrors have lattice constants of 430 nm, 435 nm, and 440 nm, exhibit a baseline reflectance of >0.5, originating from the photonic bandgap, in the broadband wavelength regions of 1526–1576 nm, 1532–1585 nm, and 1548–beyond 1600 nm, respectively. The dashed vertical and solid horizontal lines indicate the center wavelength and width of the simulated photonic bandgap, demonstrating excellent agreement with the experimental results. As shown in the Fig. 4(c), the demonstrated PhC FP micro-resonators support high-performance resonant modes with an intrinsic Q-factor on the order of $10^6$ across the S, C, and L band.

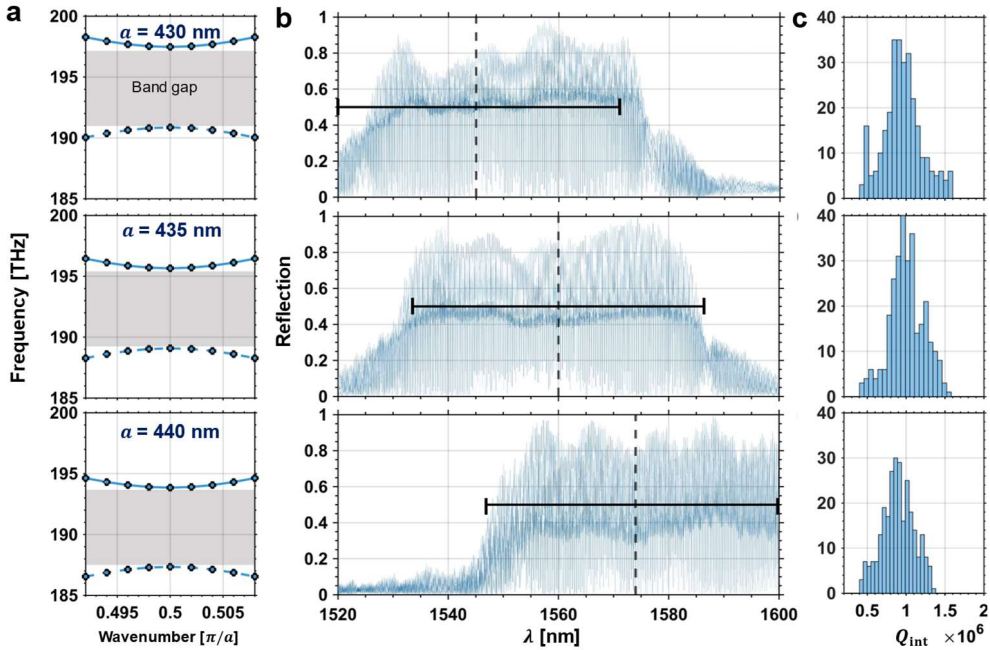

Fig. 4 (a) Simulated photonic band structures around the one-dimensional Brillouin zone boundary for PhC lattice periods of 430 nm, 435 nm, and 440 nm. (b) Experientially measured reflection spectra of PhC FP micro-resonators with corresponding PhC lattice constants. The dashed vertical and solid horizontal lines indicate the center wavelength and width of the simulated photonic bandgap, respectively. The cavity length is fixed with 4mm. (c) Statistical distribution of intrinsic Q-factors.

## 6. Systematic control of coupling strength

The reflectance and coupling strength of the PhC mirror can be systematically engineered. Figure 5(a) illustrates the schematic of the PhC mirror, which consists of a central region with $m_C$ identical unit cells, sandwiched between two apodization regions, each containing $m_T$ unit cells that gradually vary. The apodization regions minimize the mode mismatch between the propagating wave in the coupled waveguide or cavity and the PhC Bloch wave, thereby suppressing undesired optical losses into free space. With $m_T$ fixed, varying $m_C$ effectively controls the coupling strength between the input waveguide and the micro-cavity through the PhC mirror. In this work, $m_C$ varies from 1 to 20, while $m_T$ is set to 30.

We numerically calculate the coupling strength of the PhC coupler using full three-dimensional finite-difference time-domain (FDTD) simulations (Tidy3D, Flexcompute). Figure 5(b) presents the simulated distributions of the dominant components of the electric field and Poynting vector, $E_y$ and $S_x$, respectively. The fundamental TE$_{00}$ mode is launched at a position in the input waveguide on the left, as indicated by the orange arrow, and is incident on the PhC coupler, which reflects part of the incident wave back into the waveguide and transmits the rest into the cavity on the right. The reflectance and transmittance of the PhC coupler are determined by monitoring the flux of the Poynting vector. The relationship between the coupling Q-factor ($Q_{cpl}$) and transmittance ($T$) is given by: $\boldsymbol{Q_{cpl}} = -\frac{\omega_0 t_R}{\ln(1-T)}$, where $t_R$ and $\omega_0$ is the round-trip time and frequency of the resonant mode of the PhC FP micro-resonator. The round-trip time is determined by the group index and cavity length in the PhC FP micro-resonator.

We compare the theoretical and experimental values of $Q_{cpl}$ for three different couplers with $m_C$ = 1, 10, and 20, which show good agreement with each other, as shown in Fig. 5(c). Here, the cavity length of the PhC FP micro-resonator is fixed at 4 mm. As $m_C$ increases, the coupling Q-factor increases due to the reduced transmission through the PhC coupler. The broadband performance of the photonic bandgap and the wavelength at which the maximum $Q_{cpl}$ occurs remain unchanged. While directional couplers for conventional waveguides and micro-ring resonators suffer from coupling strength variations that are highly sensitive to the wavelength of light,[10,14] the PhC coupler enables wavelength-independent control of $Q_{cpl}$ across its photonic bandgap. Despite its small footprint of only ~100 μm (Fig. 1(c)), the PhC coupler allows systematic tuning of the coupling strength across the under-, critical-, and over-coupling regimes, complementing the dependence of the intrinsic Q-factor on the cavity length in the PhC FP micro-resonator platform. It is important to note that the design of the PhC coupler is decoupled from the resonator size, enabling independent control of the $Q_{cpl}$. This contrasts with directional couplers in ring resonator systems, where the coupler size increases with the cavity size. We note that the PhC coupler is particularly advantageous for applications that require over-coupling (lower $Q_{cpl}$ values), which can be easily accomplished by reducing the number of unit cells.

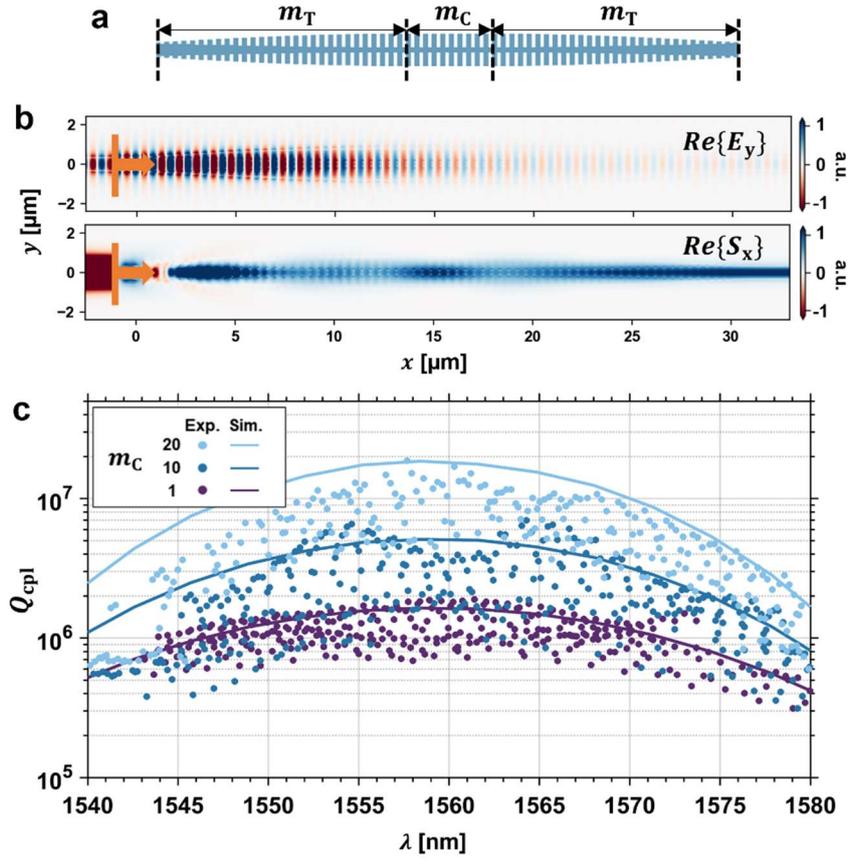

**Fig. 5** (a) Schematic of the one-dimensional PhC mirror, consisting of a central region with $m_C$ identical unit cells, sandwiched between two apodization regions with $m_T$ unit cells. (b) Full three-dimensional FDTD simulation results of the electric field ($E_y$) and Poynting vector ($S_x$). The fundamental $TE_{00}$ mode (orange arrow) is launched at the input waveguide (left, orange arrow) and interacts with the PhC mirror. (c) Comparison of calculated and measured coupling Q-factors ($Q_{cpl}$) for three different PhC couplers ($m_C$ = 1, 10, and 20).

# 7. Conclusion

In this work, we demonstrate a high-Q TFLN PhC FP micro-resonator and perform a comprehensive analysis of its scaling behavior, spectral tunability, and coupling strength control. Enabled by low propagation loss and minimized interfacial loss at the PhC reflector and coupler, the standing-wave resonant modes exhibit intrinsic Q-factors of in the order of $10^6$ over the spectral range of the photonic bandgap, setting a new performance benchmark for one-dimensional FP resonators in TFLN. The intrinsic Q-factor and the FSR are systematically scaled with the FP cavity length, in contrast to conventional traveling-wave resonators, which inherently suffer from the size-dependent dispersion and group index variation. The coupling strength of the PhC coupler can be consistently controlled by engineering the number of its unit cells. Moreover, the photonic bandgap can be tuned across the S-, C-, and L-bands in the telecommunication wavelength range, without degradation of resonator performance. Locking the spectral range of high-Q resonant modes within the photonic bandgap offers strong potential to suppress parasitic nonlinearities, such as Raman scattering, and to enable low-noise, high-efficiency second-order nonlinear applications in TFLN. Notably, the one-dimensional geometry of the PhC FP micro-resonator facilitates significantly higher integration density of photonic components on a chip, compared to conventional two-dimensional traveling-wave micro-resonators. We believe that, given these demonstrated advantages and innovations, the high-Q TFLN PhC FP micro-resonator will open new avenues in photonic integrated circuits for nonlinear optics and quantum applications.

# 8. Method

### A. Design method for low-loss photonic crystal reflectors

This section describes the design methodology for a low-loss PhC reflector, as shown in Fig. 6. The procedure consists of three main steps: First, the PhC period ($a$) is chosen to align the center wavelength ($\lambda_c$) of the photonic bandgap with the target operating wavelength. Band diagram simulations in Fig. 6(a) show that a period of 435 nm positions the bandgap center near 1560 nm as a representative example. Second, the initial waveguide width ($W_1$) is determined using the Bragg condition $2n_{\text{eff}} a = \lambda_c$, where $n_{\text{eff}}$ is the effective index of the propagation mode in the waveguide of width $W_1$. Figure 6(b) presents the combinations of waveguide width and PhC period that satisfy the Bragg condition, for instance, $W_1$ = 800 nm when $a$ = 435 nm. Third, the reflector performance is evaluated using full 3D FDTD simulations. Figure 6(c) plots the reflectance ($R$) and the sum of the reflectance and transmittance ($R + T$) on a logarithmic scale, where both $R$ and $T$ are calculated with respect to the guided waveguide mode. Accordingly, $1 - (R + T)$ corresponds to radiation leakage into free space, primarily due to scattering losses in the PhC structure. The optimized design with $W_1$ = 800 nm exhibits low-loss reflection over the photonic bandgap region centered at 1560 nm, whereas the non-optimized case with $W_1$ = 1000 nm introduces additional loss at the short wavelength due to the leaky mode excitation. The corresponding field profiles in Fig. 6(d) reveal excitation of leaky slab mode. Although the simulation validates 800 nm as the optimal $W_1$, the fabricated device in our experiment used $W_1$ = 700 nm, due to fabrication process variations.

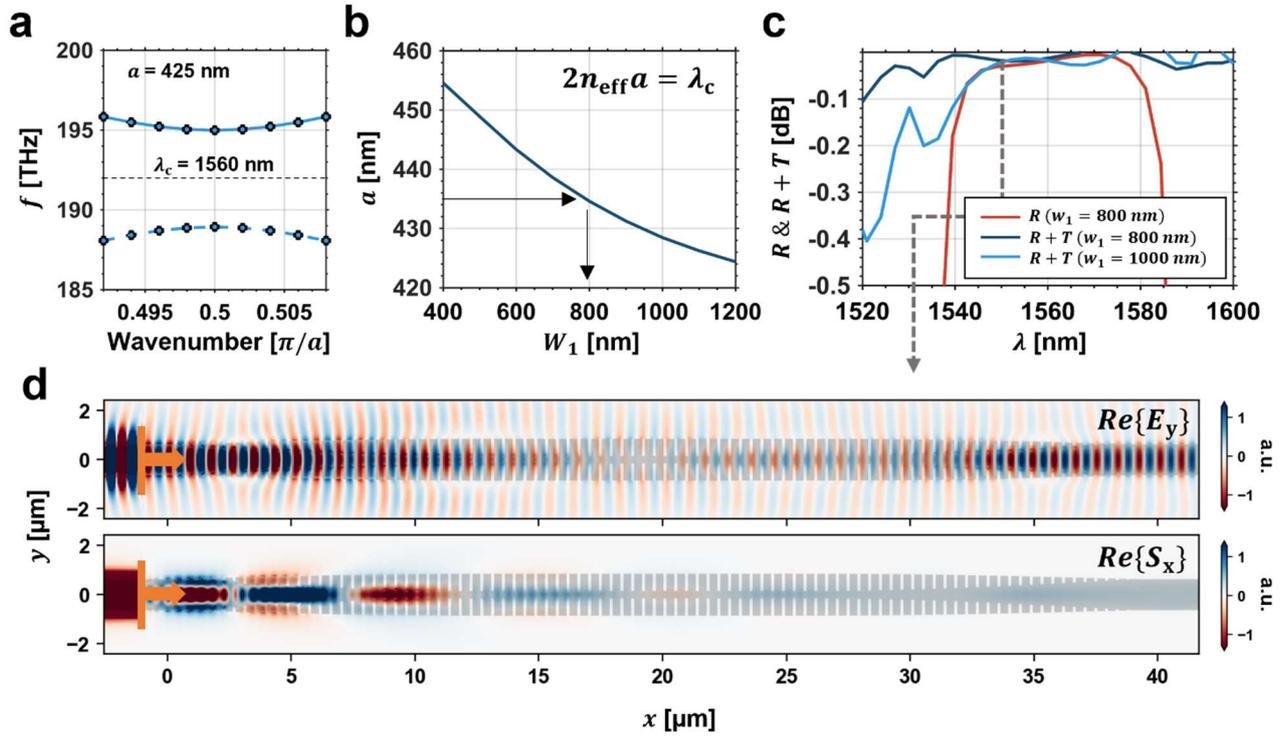

**Fig. 6** Design methodology of a low-loss PhC reflector. (a) Photonic band simulation for a PhC with a period of 435 nm, reflecting the geometrical parameters in Section 2. (b) Bragg condition for $\lambda_c$ = 1560 nm, plotted as a function of waveguide width and PhC period. (c) Simulated reflectance (R) and the sum of reflectance and transmittance (R + T) for optimized ($W_1$ = 800 nm) and non-optimized ($W_1$ = 1000 nm) designs. (d) Distribution of the electric field $Re(E_y)$ and Poynting vector $Re(S_x)$ for $W_1$ = 1000 nm at the wavelength of 1550 nm.

## 9. Acknowledgments

M.K.S and H.L acknowledge support by KAIST Cross-Generation Collaborative Lab project. the Institute for Information, Communication Technology Promotion of Korea (2021-0-00552), the National Research Foundation of Korea (RS-2024-00350185, RS-2024-00408271), the National Research Council of Science and Technology (21031-200).

## 10. Author declarations

**Conflict of Interest**

The authors have no conflicts to disclose.

**Author Contributions**

**Hyeon Hwang:** Conceptualization (lead); Investigation (lead); Data curation (lead); Formal Analysis (lead); Visualization (lead); Writing/Original Draft Preparation (lead); Writing/Review & Editing (equal); **Seokjoo Go:** Data curation (supporting); Resources (supporting); Writing/Review & Editing (supporting); **Guhwan Kim:** Investigation (supporting); **Hong-Seok Kim**: Investigation (supporting); **Kiwon Moon:** Investigation (supporting); Resources (supporting); **Jung Jin Ju:** Investigation (supporting); Resources (supporting); **Hansuek Lee:** Conceptualization (equal);   Funding Acquisition (equal); Supervision (equal); Writing/Review & Editing (equal); **Min-Kyo Seo:** Conceptualization (equal);   Funding Acquisition (equal); Project Administration (equal); Supervision (equal); Writing/Original Draft Preparation (equal); Writing/Review & Editing (equal);

## 11. Data availability

The data that support the findings of this study are available from the corresponding author upon reasonable request.

## 12. References

[1] G. Chen, N. Li, J. Da Ng, H.L. Lin, Y. Zhou, Y.H. Fu, L.Y.T. Lee, Y. Yu, A.Q. Liu, and A.J. Danner, "Advances in lithium niobate photonics: development status and perspectives," Advanced Photonics **4**(3), (2022).

[2] E.L. Wooten, K.M. Kissa, A. Yi-Yan, E.J. Murphy, D.A. Lafaw, P.F. Hallemeier, D. Maack, D.V. Attanasio, D.J. Fritz, G.J. McBrien, and D.E. Bossi, "A review of lithium niobate modulators for fiber-optic communications systems," IEEE Journal of Selected Topics in Quantum Electronics **6**(1), 69–82 (2000).

[3] A. Boes, L. Chang, C. Langrock, M. Yu, M. Zhang, Q. Lin, M. Lončar, M. Fejer, J. Bowers, and A. Mitchell, "Lithium niobate photonics: Unlocking the electromagnetic spectrum," Science (1979) **379**(6627), (2023).

[4] D. Zhu, L. Shao, M. Yu, R. Cheng, B. Desiatov, C.J. Xin, Y. Hu, J. Holzgrafe, S. Ghosh, A. Shams-Ansari, E. Puma, N. Sinclair, C. Reimer, M. Zhang, and M. Lončar, "Integrated photonics on thin-film lithium niobate," Adv Opt Photonics **13**(2), 242 (2021).

[5] K. Zhang, W. Sun, Y. Chen, H. Feng, Y. Zhang, Z. Chen, and C. Wang, "A power-efficient integrated lithium niobate electro-optic comb generator," Commun Phys **6**(1), (2023).

[6] C. Wang, M. Zhang, B. Stern, M. Lipson, and M. Lončar, "Nanophotonic lithium niobate electro-optic modulators," Opt Express **26**(2), 1547 (2018).

[7] M. Li, J. Ling, Y. He, U.A. Javid, S. Xue, and Q. Lin, "Lithium niobate photonic-crystal electro-optic modulator," Nat Commun **11**(1), (2020).

[8] M. Zhang, C. Wang, P. Kharel, D. Zhu, and M. Lončar, "Integrated lithium niobate electro-optic modulators: when performance meets scalability," Optica **8**(5), 652 (2021).

[9] C. Wang, M. Zhang, X. Chen, M. Bertrand, A. Shams-Ansari, S. Chandrasekhar, P. Winzer, and M. Lončar, "Integrated lithium niobate electro-optic modulators operating at CMOS-compatible voltages," Nature **562**(7725), 101–104 (2018).

[10] H. Hwang, M.R. Nurrahman, H. Heo, K. Ko, K. Moon, J.J. Ju, S.-W. Han, H. Jung, H. Lee, and M.-K. Seo, "Hyperband electro-optic modulator based on a two-pulley coupled lithium niobate racetrack resonator," Opt Lett **49**(3), 658 (2024).

[11] C. Wang, M. Zhang, M. Yu, R. Zhu, H. Hu, and M. Loncar, "Monolithic lithium niobate photonic circuits for Kerr frequency comb generation and modulation," Nat Commun **10**(1), (2019).

[12] T.H. Wu, L. Ledezma, C. Fredrick, P. Sekhar, R. Sekine, Q. Guo, R.M. Briggs, A. Marandi, and S.A. Diddams, "Visible-to-ultraviolet frequency comb generation in lithium niobate nanophotonic waveguides," Nat Photonics **18**(3), 218–223 (2024).

[13] M. Zhang, B. Buscaino, C. Wang, A. Shams-Ansari, C. Reimer, R. Zhu, J.M. Kahn, and M. Lončar, "Broadband electro-optic frequency comb generation in a lithium niobate microring resonator," Nature **568**(7752), 373–377 (2019).

[14] Y. Song, Y. Hu, X. Zhu, K. Yang, and M. Lončar, "Octave-spanning Kerr soliton frequency combs in dispersion- and dissipation-engineered lithium niobate microresonators," Light Sci Appl **13**(1), (2024).

[15] P.I. Sund, E. Lomonte, S. Paesani, Y. Wang, J. Carolan, N. Bart, A.D. Wieck, A. Ludwig, L. Midolo, W.H.P. Pernice, P. Lodahl, and F. Lenzini, "High-speed thin-film lithium niobate quantum processor driven by a solid-state quantum emitter," Sci Adv **9**(19), (2023).

[16] H.S. Stokowski, T.P. McKenna, T. Park, A.Y. Hwang, D.J. Dean, O.T. Celik, V. Ansari, M.M. Fejer, and A.H. Safavi-Naeini, "Integrated quantum optical phase sensor in thin film lithium niobate," Nat Commun **14**(1), (2023).

[17] X. Wang, X. Jiao, B. Wang, Y. Liu, X.P. Xie, M.Y. Zheng, Q. Zhang, and J.W. Pan, "Quantum frequency conversion and single-photon detection with lithium niobate nanophotonic chips," Npj Quantum Inf **9**(1), (2023).

[18] S. Saravi, T. Pertsch, and F. Setzpfandt, "Lithium Niobate on Insulator: An Emerging Platform for Integrated Quantum Photonics," Adv Opt Mater **9**(22), (2021).

[19] J. Zhao, C. Ma, M. Rüsing, and S. Mookherjea, "High Quality Entangled Photon Pair Generation in Periodically Poled Thin-Film Lithium Niobate Waveguides," Phys Rev Lett **124**(16), (2020).

[20] H. Hwang, woojin noh, M. Nurrahman, G. Kim, K. Moon, J. Ju, H. Lee, and M.-K. Seo, "Spontaneous parametric down-conversion photon pair generation in small footprint X-cut periodically poled lithium niobate micro-resonator," Opt Lett, (2024).

[21] R. Luo, H. Jiang, S. Rogers, H. Liang, Y. He, and Q. Lin, "On-chip second-harmonic generation and broadband parametric down-conversion in a lithium niobate microresonator," Opt Express **25**(20), 24531 (2017).

[22] J. Lu, M. Li, C.-L. Zou, A. Al Sayem, and H.X. Tang, "Toward 1% single-photon anharmonicity with periodically poled lithium niobate microring resonators," Optica **7**(12), 1654 (2020).

[23] J. Lu, J.B. Surya, X. Liu, A.W. Bruch, Z. Gong, Y. Xu, and H.X. Tang, "Periodically poled thin-film lithium niobate microring resonators with a second-harmonic generation efficiency of 250,000%/W," Optica **6**(12), 1455 (2019).

[24] M. Zhang, C. Wang, R. Cheng, A. Shams-Ansari, and M. Lončar, "Monolithic ultra-high-Q lithium niobate microring resonator," Optica **4**(12), 1536 (2017).

[25] A. Pan, C. Hu, C. Zeng, and J. Xia, "Fundamental mode hybridization in a thin film lithium niobate ridge waveguide," Opt Express **27**(24), 35659 (2019).

[26] X. Zheng, Z. Lin, Q. Huang, and S. He, "Elimination of the fundamental mode hybridization on an x-cut lithium-niobate-on-insulator by using a densely packed bent waveguide array," Appl Opt **62**(21), 5765 (2023).

[27] K. Chen, Y. Liu, X. Xing, H. Wang, T. Gao, W. Ma, F. Zhao, L. Wu, C. Guo, and L. Liu, "Polarization Management on Anisotropic Thin-Film Lithium Niobate Platform," Adv Photonics Res, (2025).

[28] M. Yu, Y. Okawachi, R. Cheng, C. Wang, M. Zhang, A.L. Gaeta, and M. Lončar, "Raman lasing and soliton mode-locking in lithium niobate microresonators," Light Sci Appl **9**(1), (2020).

[29] H. Fan, Z. Ma, J. Chen, Z. Li, C. Tang, Y.M. Sua, and Y. Huang, "Photon conversion in thin-film lithium niobate nanowaveguides: a noise analysis," Journal of the Optical Society of America B **38**(7), 2172 (2021).

[30] T. Wildi, M.A. Gaafar, T. Voumard, M. Ludwig, and T. Herr, "Dissipative Kerr solitons in integrated Fabry–Perot microresonators," Optica **10**(6), 650 (2023).

[31] Y. Zhang, S. Veilleux, and M. Dagenais, "Fabry-Perot Bragg grating nanoresonator with ultrahigh intrinsic Q based on low-loss silicon nitride," Opt Express **31**(21), 34688 (2023).

[32] S. Xie, Y. Zhang, Y. Hu, S. Veilleux, and M. Dagenais, "On-Chip Fabry-Perot Bragg Grating Cavity Enhanced Four-Wave Mixing," ACS Photonics **7**(4), 1009–1015 (2020).

[33] C. Ye, Y. Liu, C. Kim, K. Yvind, and M. Pu, "Singly-Resonant Four-Wave Mixing Based onanOn-Chip Fabry-Perot Bragg Grating Cavity," in *2023 IEEE Photonics Conference, IPC 2023 - Proceedings*, (Institute of Electrical and Electronics Engineers Inc., 2023).

[34] A. Nardi, A. Davydova, N. Kuznetsov, M.H. Anderson, C. Möhl, J. Riemensberger, T.J. Kippenberg, and P. Seidler, "Integrated chirped photonic-crystal cavities in gallium phosphide for broadband soliton generation," Optica **11**(10), 1454 (2024).

[35] A.E. Ulanov, T. Wildi, U. Bhatnagar, and T. Herr, "Laser diode self-injection locking to an integrated high-Q Fabry–Perot microresonator," Opt Lett **49**(21), 6261 (2024).

[36] H. Hwang, H. Heo, K. Ko, M.R. Nurrahman, K. Moon, J.J. Ju, S.-W. Han, H. Jung, H. Lee, and M.-K. Seo, "Electro-optic control of the external coupling strength of a high-quality-factor lithium niobate micro-resonator," Opt Lett **47**(23), 6149 (2022).

[37] H. Hwang, K. Ko, M.R. Nurrahman, K. Moon, J.J. Ju, S.W. Han, H. Jung, M.K. Seo, and H. Lee, "A wide-spectrum mid-infrared electro-optic intensity modulator employing a two-point coupled lithium niobate racetrack resonator," APL Photonics **10**(1), (2025).

[38] D.R. Carlson, D.D. Hickstein, W. Zhang, A.J. Metcalf, F. Quinlan, S.A. Diddams, and S.B. Papp, "Ultrafast electro-optic light with subcycle control," Science (1979) **361**(6409), 1358–1363 (2018).

[39] P. Del'Haye, A. Schliesser, O. Arcizet, T. Wilken, R. Holzwarth, and T.J. Kippenberg, "Optical frequency comb

generation from a monolithic microresonator," Nature **450**(7173), 1214–1217 (2007).

[40] J. Li, H. Lee, T. Chen, and K.J. Vahala, "Characterization of a high coherence, Brillouin microcavity laser on silicon," Opt Express **20**(18), 20170 (2012).

[41] K. Ko, D. Suk, D. Kim, S. Park, B. Sen, D.-G. Kim, Y. Wang, S. Dai, X. Wang, R. Wang, B.J. Chun, K.-H. Ko, P.T. Rakich, D.-Y. Choi, and H. Lee, "A mid-infrared Brillouin laser using ultra-high-Q on-chip resonators," Nat Commun **16**(1), 2707 (2025).

[42] M. Xu, M. He, Y. Zhu, L. Liu, L. Chen, S. Yu, and X. Cai, "Integrated thin film lithium niobate Fabry–Perot modulator [Invited]," Chinese Optics Letters **19**(6), 060003 (2021).

[43] J. Cheng, D. Gao, J. Dong, and X. Zhang, "Efficient second harmonic generation in a high-Q Fabry-Perot microresonator on x-cut thin film lithium niobate," Opt Express **32**(7), 12118 (2024).

[44] J. Kellner, A. Sabatti, A. Meader, and R. Grange, "Low threshold integrated optical parametric oscillator with a compact Bragg resonator," (2025).

[45] X. Ji, J. Liu, J. He, R.N. Wang, Z. Qiu, J. Riemensberger, and T.J. Kippenberg, "Compact, spatial-mode-interaction-free, ultralow-loss, nonlinear photonic integrated circuits," Commun Phys **5**(1), (2022).